\documentclass[prl,twocolumn,preprintnumbers,amsmath,amssymb,floats,showpacs]{revtex4}
\usepackage{tipa}

\usepackage{graphicx}

\begin{document}

\title{A novel mechanism of charge density wave in a transition metal dichalcogenide}
\author{D. W. Shen$^1$, B. P. Xie$^1$, J. F. Zhao$^1$, L. X. Yang$^1$,
 L. Fang$^2$,
 J. Shi$^3$,
 R. H. He$^4$, D. H. Lu$^4$,
 H. H. Wen$^{2}$, and
 D.L. Feng$^{1}$} \email{dlfeng@fudan.edu.cn}
 \affiliation{$^1$Department of Physics, Applied Surface Physics State Key Laboratory,
 Fudan University, Shanghai 200433, China}
 \affiliation{$^2$National Lab for Superconductivity, Institute of Physics and National
 Lab for Condensed Matter Physics, Chinese Academy of Sciences, P.O. Box 603, Beijing 100080, P. R. China}
 \affiliation{$^3$School of Physics, Wuhan University, Wuhan, 430072, P. R. China}
 \affiliation{$^4$Department of Applied Physics and Stanford Synchrotron Radiation Laboratory,
  Stanford University, Stanford, CA 94305, USA}

\date{\today}

\begin{abstract}
Charge density wave, or CDW, is usually associated with Fermi
surfaces nesting. We here report a new CDW mechanism discovered in
a 2$H$-structured transition metal dichalcogenide, where the two
essential ingredients of CDW are realized in very anomalous ways
due to the strong-coupling nature of the electronic structure.
Namely, the CDW gap is only partially open, and charge density
wavevector match is fulfilled through participation of states of
the large Fermi patch, while the straight FS sections have
secondary or negligible contributions.
\end{abstract}


\maketitle

It has been a standard textbook-example that charge density wave
(CDW), one of the main forms of ordering in solid, is mostly
associated with nesting Fermi surface (FS) sections. In charge
ordered materials ranging from one-dimensional (1D)
(TaSe$_{4}$)$_{2}$I and blue bronze\cite{DardelPRL,DardelEPL} to
two-dimensional (2D) manganites, and from surface reconstruction
in weak correlated metals to checker board pattern of strongly
correlated high temperature superconductors\cite{ShenKM,Nakagawa},
the charge fluctuations associated with the ordering wave vector
scatter the electrons between two nested FS sections and
effectively drive the system into an ordered ground state.
However, this classical picture failed in the very first 2D CDW
compound discovered in 1974, \emph{i.e.} the transition metal
dichalcogenides (TMD's)\cite{WilsonAP}. The 2$H$-structured TMD's
have a hexagonal lattice structure, and in its CDW phase, a
$3\times3$ superlattice forms\cite{WilsonAP,Moncton1}. It was
found that the ordering wavevectors do not match the nested FS
sections, and generally no CDW energy gap was observed at the
FS\cite{RossnagelPRB,TonjesPRB}. After decades of continuous
effort, the origin of CDW for the $2H$-structured TMD's has been a
long standing mystery. As a result, the subtle details of the
competition and coexistence of CDW and superconductivity in
TMD's\cite{Castro} remain to be revealed\cite{shinScience}.

In this letter, we studied the electronic origin of the CDW in a
2$H$-TMD, $2H$-Na$_{x}$TaS$_{2}$, by angle-resolved photoemission
spectroscopy (ARPES). The CDW mechanism in this material was
discovered after the revelation of the following exotic properties.
 i) The electronic structure exhibits strong coupling
nature, with finite density of states at the Fermi energy($E_F$)
over almost the entire Brillouin zone(BZ), forming so-called Fermi
patches. ii) In the CDW state, only a fraction of the states at
$E_F$ is gapped. iii) The density of states near $E_F$ directly
correlates with the ordering strength. iv) Fermi patch, instead of
Fermi surface, is relevant for CDW. We show that this new
``Fermi-patch mechanism" for CDW is rooted in the strong coupling
nature of the electronic structure and it may be a general theme
of ordering in the strong coupling regime of various models, and
applicable to systems with similar electronic structure.

For the systematic studies of the electronic structure in a
2$H$-TMD compound, 2$H$-Na$_{x}$TaS$_{2}$ with 2$\%$, 5$\%$ and
10$\%$ Na concentration were synthesized with CDW transition
temperature T$_{CDW}$'s are 68K, 65K and 0K respectively. The
samples are labelled as CDW$^{68K}$, CDW$^{65K}$, and CDW$^{0K}$
hereafter. The corresponding superconducting transition
temperatures are 1.0K, 2.3K and 4.0K respectively, a manifestation
of the competition between CDW and superconductivity in this
system. The data were mainly collected using 21.2eV Helium-I line
of a discharge lamp combined with a Scienta R4000 analyzer, and
partial measurements were carried out on beam line 5-4  of SSRL.
The overall energy resolution is 8 $meV$, and the angular
resolution is 0.3 degrees.

\begin{figure*}[t]
\includegraphics[width=15cm]{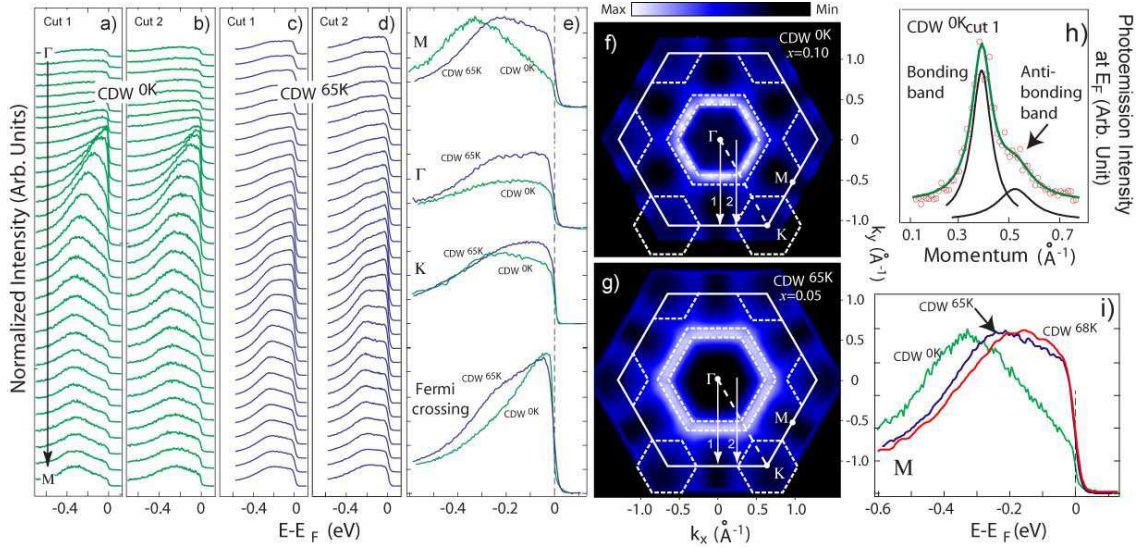}
\caption{Typical ARPES spectra for \textbf{a-b}, CDW$^{0K}$ at
T=15K and \textbf{c-d}, CDW$^{65K}$ at T=95K respectively along
the marked cuts in the Brillouin zone in panels \textbf{f} and
\textbf{g}. \textbf{e}, Comparison of photoemission spectra for
different $2H$-Na$_{x}$TaS$_{2}$'s at high symmetry points
\textbf{M}, \textbf{$\Gamma$}, \textbf{K} and at the Fermi
crossing of cut1. \textbf{f} and \textbf{g}, The photoemission
intensity integrated within 10meV of $E_F$ is shown for
 CDW$^{0K}$ and  CDW$^{65K}$ respectively. (Both data were taken at 15K and the image was
6-fold symmetrized.) The Fermi surfaces are marked by dashed
lines, where the antibonding and bonding bands could be resolved
for the $\Gamma$ pockets in the momentum distribution curve (one
example is shown in \textbf{h} for CDW$^{0K}$). } \label{EDC}
\vspace*{-0.5cm}
\end{figure*}

Photoemission spectra taken on samples with (CDW$^{65K}$) and
without CDW (CDW$^{0K}$) are compared in
Fig.\,\ref{EDC}\textbf{a-d}. In both cases, the spectral
lineshapes are remarkably broad, and no quasiparticle peaks in the
conventional sense are observed. The large linewidth is of the
same order as the dispersion, which clearly indicates the
incoherent nature of the spectrum and the system is in the
\textit{strong coupling} regime. Taking the normalized spectra at
\textbf{M} as reference\cite{exp1}, spectra at \textbf{$\Gamma$},
$\textbf{K}$, and the Fermi crossing of the
\textbf{$\Gamma$}-\textbf{M} cut are compared in
Fig.\,\ref{EDC}\textbf{e}. The difference between CDW$^{0K}$ and
CDW$^{65K}$ is striking. Although they have similar density of
states at the Fermi crossing area, for momentum regions away from
the FS, CDW$^{65K}$ has much stronger spectral weight than
CDW$^{0K}$, no matter whether it is inside the occupied region
(\textbf{M} point), or in the unoccupied region (\textbf{$\Gamma$}
and \textbf{K} points) in the band structure
calculations\cite{Guo}. Particularly, one finds that even when the
spectral centroid is well below $E_F$, the finite residual weight
at $E_F$ beyond background exists around \textbf{M}. We note that
the CDW$^{0K}$ sample has higher Na doping than CDW$^{65K}$, yet
its lineshape is generally sharper.  Therefore, disorder effects
induced by the dopants should be negligible. The residual weight
observed  near $E_F$  thus should be associated with the intrinsic
strong coupling nature of the system, which (within several
$k_{B}T_{CDW}$ around $E_F$) shows a monotonic correlation with
T$_{CDW}$ in Fig.\,\ref{EDC}\textbf{i}.

Although the spectra are broad, Fermi surfaces could still be
defined as the local maximum of the spectral weight at
$E_F$\cite{KipPRL}, which are plotted in
Fig.\,\ref{EDC}\textbf{f}-\textbf{g} for CDW$^{0K}$ and
CDW$^{65K}$ respectively. Three FS pockets can be identified
through the momentum distribution curve analysis: two hole pockets
around \textbf{$\Gamma$}, as exemplified in
Fig.\,\ref{EDC}\textbf{h}, and one hole pocket around \textbf{K}.
Bilayer band splitting of the two TaS$_{2}$ layers in a unit cell
manifests itself as the inner and outer Gamma pockets, while the
splitting of the  \textbf{K} pockets is indistinguishable. The FS
volumes are evaluated to be $1.05\pm0.01$ and $1.11\pm0.01$
electrons per layer for the 5$\%$ and 10$\%$ Na doped
Na$_{x}$TaS$_{2}$ samples respectively, consistent with the
nominal dopant concentrations. This resembles the high temperature
superconductors, where the Fermi surfaces defined in this
conventional way are consistent with the band structure
calculations and follow the Luttinger sum rule quite
well\cite{dinghongPRL1997}, yet there are large Fermi patches near
the antinodal regions\cite{FurukawaPRL}. In CDW$^{65K}$ case,
almost the entire BZ appear to be one gigantic Fermi patch.
However, so far, searches for the CDW mechanism are mostly
centered on the Fermi surfaces, not the Fermi patch.

\begin{figure}[t]
\includegraphics[width=9cm]{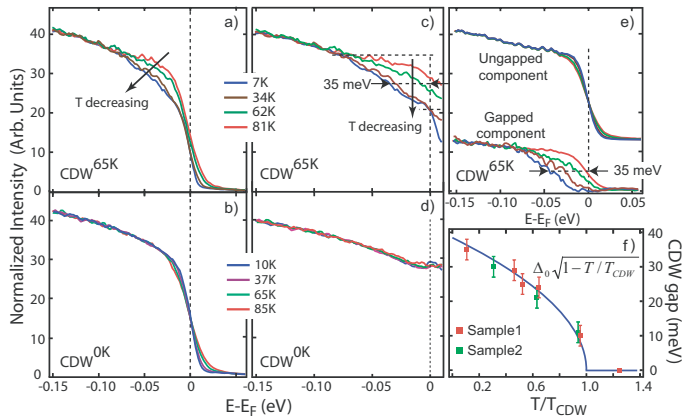}
\caption{The CDW gap measurements of 2$H$-Na$_{x}$TaS$_{2}$. ARPES
spectra taken at \textbf{M} for different temperatures for
\textbf{a}, CDW$^{65K}$ and \textbf{b}, CDW$^{0K}$. \textbf{c-d},
The spectra in \textbf{a} and \textbf{b} divided by the resolution
convoluted Fermi-Dirac distribution at the corresponding
temperatures. \textbf{e}, Each spectrum in \textbf{a} is
decomposed into a gapped component and an ungapped component.
\textbf{f}, The temperature dependence of the CDW gap. The solid
line is the fit to a mean field formula,
$\Delta_{0}\sqrt{1-\frac{T}{T_{CDW}}}$.} \label{decomp}
\vspace*{-0.5cm}
\end{figure}

The differences between CDW$^{65K}$ and CDW$^{0K}$ spectra in the
Fermi patch region(Fig.\,\ref{EDC}\textbf{e}) are intriguing.
Fig.\,\ref{decomp}\textbf{a-b} show spectra taken from \textbf{M}
at different temperatures for CDW$^{65K}$ and CDW$^{0K}$
respectively. While the CDW$^{0K}$ spectra simply exhibit a clear
Fermi crossing and thermal broadening, the CDW$^{65K}$ spectra
appear very anomalous. Take the spectrum at 7K as an example,
while the upper part of the spectrum is suppressed to higher
binding energies, the middle point of the leading edge of the
lower part still matches $E_F$. There is an apparent turning point
between these two parts of the spectrum. By dividing the spectra
with the corresponding finite temperature Fermi-Dirac distribution
functions, in Fig.\,\ref{decomp}\textbf{c}, one clearly observes
that about 29\% of the spectral weight at $E_F$ has been
suppressed for CDW$^{65K}$. An energy gap has opened on part of
the states here, which is estimated to be $\sim$35$meV$ based on
the middle point of the leading edge. Contrastively, there is no
sign of gap opening for CDW$^{0K}$ (Fig.\,\ref{decomp}\textbf{d}).

It was recently suggested by Barnett and
coworkers\cite{BarnettPRL} that a 2$H$-TMD system is decoupled
into three sub-lattices. While one of the sub-lattices is
undistorted and gapless below T$_{CDW}$, the other two are gapped
at the FS. For CDW$^{65K}$, similar behavior is observed except
that gap does not open at Fermi Surface and just about 1/3 of the
spectrum is gapped. \textit{Phenomenologically}, one can decompose
the spectra into gapped and ungapped components. The ungapped
component is simulated as
$$A_{u}(k,\omega,T)= \alpha
A(k,\omega,85K)\frac{f(\omega,T)}{f(\omega,85K)}$$, where
$A(k,\omega,T)$ is the spectral function,  $f(\omega,T)$ is the
resolution convoluted finite temperature Fermi-Dirac distribution
function, and $\alpha=0.71$. The gapped components
$$A_{g}(k,\omega,T)=A(k,\omega,T)-A_{u}(k,\omega,T)$$.
The decomposition of spectra in Fig.\,\ref{decomp}\textbf{a} is
shown in Fig.\,\ref{decomp}\textbf{e}, where the ungapped
components exhibit the same behavior as in the CDW$^{0K}$
spectrum, and the gapped components clearly reveal a clean energy
gap of about 35meV at 7K. We emphasize that although the
decomposition method is adopted hereafter, all the qualitative
results can be obtained through the conventional method in
Fig.\,\ref{decomp}\textbf{c} as well. The temperature evolution of
the gap is shown in Fig.\,\ref{decomp}\textbf{f} for two different
CDW$^{65K}$ samples. Interestingly, the gap does not saturate at
low temperature, and it can be fitted very well to
$\Delta_{0}\sqrt{1-\frac{T}{T_{CDW}}}$, which is the mean field
theory form at temperature close to T$_{CDW}$\cite{Gruner}. Here
the fitted
$\Delta_{0}$$\sim$39$meV$$\sim$6$\emph{k}_{B}$$\cdot$T$_{CDW}$.

\begin{figure}[t]
\includegraphics[width=9cm]{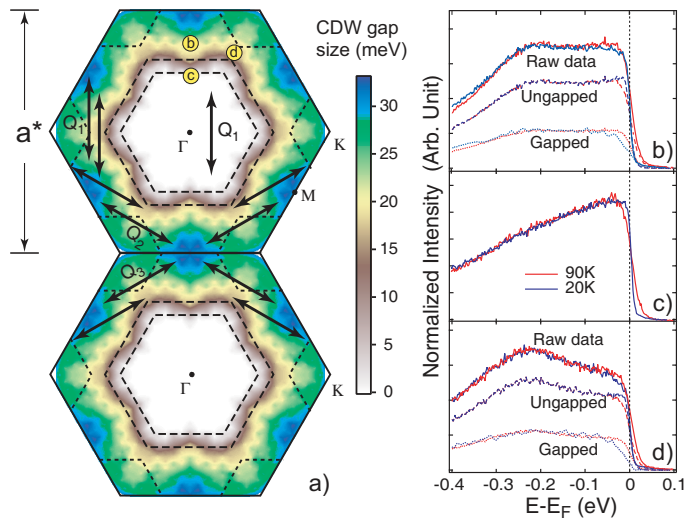}
\caption{The gap map over the Brillouin zone. \textbf{a}, The
false color plot of the CDW gap in the first Brillouin zone of
CDW$^{65K}$, where the dashed lines indicate the Fermi surfaces.
States in the gapped region could be connected by the CDW
wavevectors, \textbf{Q$_{i}$} (\emph{i}=\emph{1,2,3}), as
indicated by the double-head arrows. \textbf{b-d}, comparison of
the typical spectra in normal and CDW state at various momenta
marked by the circled letters in \textbf{a}. The gapped spectra
are decomposed into ungapped (dashed lines) and gapped portion
(dotted lines).} \label{gapmapping} \vspace*{-0.5cm}
\end{figure}

Spectra at other momenta could be decomposed in the same way quite
robustly\cite{exp3}, and the CDW gap was mapped over the entire
Brillouin zone for CDW$^{65K}$ (Fig.\,\ref{gapmapping}\textbf{a}).
Strikingly, finite CDW gap exists over most of the Brillouin zone.
Its maximum locates around \textbf{M}, and no gap is observed
around the inner \textbf{$\Gamma$} Fermi pocket and within.
Noticeably, the gap is finite in the \textbf{K} Fermi pockets, as
there is finite spectral weight at $E_F$. Close comparisons of
spectra at various momenta are shown in
Fig.\,\ref{gapmapping}\textbf{b-d}. In between \textbf{M} and the
Fermi crossing (Fig.\,\ref{gapmapping}\textbf{b}), the upper part
of the low temperature spectrum is overlayed with the normal state
spectrum, which is a sign of gap opening. In
Fig.\,\ref{gapmapping}\textbf{c}, for spectrum at the inner
\textbf{$\Gamma$} Fermi pocket, no sign of gap opening is
observed, consistent with previous
studies\cite{RossnagelPRB,TonjesPRB}.
Fig.\,\ref{gapmapping}\textbf{d} illustrates that a gap of 15$meV$
is observed at the saddle point of the band calculation. An
alternative mechanism was proposed for 2$H$-TMD's involving the
scattering between saddle points, which would cause a singularity
in density of states, and thus an anomaly in the dielectric
response function\cite{RicePRL}. The gap near this point has been
reported before\cite{LiuPRL}, but the distance  between these
points do not match the CDW ordering
wavevectors\cite{Straub,Rossnagel}. In the current gap map,
nothing abnormal is observed for this momentum.

\begin{figure}[t]
\includegraphics[width=9cm]{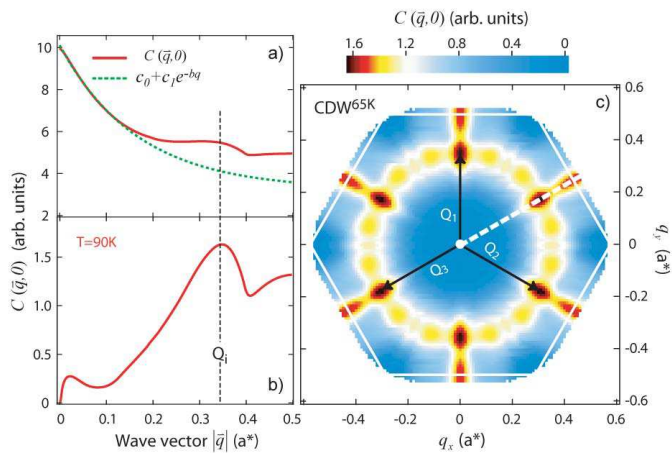}
\caption{\textbf{a} Autocorrelation of the ARPES intensity at
$E_F$ in the normal state of CDW$^{65K}$ for wavevector along the
$\Gamma$-M direction.  The dashed line is an exponential decay
plus a constant, which is deducted in \textbf{b} to highlight the
structure. \textbf{c} The partial autocorrelation in the 2D
momentum transfer space obtained in the same way as in \textbf{b}.
Repeated zone scheme is taken in the integration, and thus
$C(\vec{q},0)$ is symmetric in respect to the boundaries (white
hexagon).} \label{correlation} \vspace*{-0.5cm}
\end{figure}

The CDW in TMD's is associated with structural
transitions\cite{WilsonAP,Moncton1}, therefore, electron-phonon
interactions are also crucial in the problem. Since the low energy
electronic structure of 2$H$-Na$_{x}$TaS$_{2}$ is dominated by the
Ta 5\emph{d} electrons \cite{Guo}, among which Coulomb
interactions are usually weak, the broad ARPES lineshape would
suggests strong electron-phonon interactions. A kink in the
dispersion corresponding to the phonon energy scale was found for
2$H$-Na$_{x}$TaS$_{2}$ (not shown here) as in other 2H-TMD
compounds\cite{Valla}. In this context, the anisotropic gap
distribution might be attributed to the anisotropy of
electron-phonon couplings \cite{TD,Valla2004}, and states in the
ungapped region simply may not couple with the relevant phonons.
One critical requirement, \textit{i.e.} states in different gapped
regions need to be coupled by phonons with the CDW wavevectors,
can now be fulfilled in the gap map, as illustrated by the arrows
in Fig.\,\ref{gapmapping}\textbf{a}.
\textbf{Q$_{i}$}=\textbf{a$_{i}$}*/3,(\emph{i=1,2,3}) here are the
CDW wavevectors, \textbf{a$_{i}$}*'s being the reciprocal lattice
vectors along the three \textbf{$\Gamma$}-\textbf{M} directions.
This also explains why the size of the Fermi surfaces can vary
significantly for different 2H-TMD systems, with nearly
system-independent CDW ordering
wavevectors\cite{Straub,Rossnagel,Moncton2}.

One open question in the above picture is that charge fluctuations
with other wavevectors are also allowed, and it is not obvious why
\textbf{Q$_{i}$}'s are special. For cuprate superconductors, it
has been demonstrated that the autocorrelation of ARPES spectra,
$$C(\vec{q},\omega)\equiv\int A(\vec{k},\omega) A(\vec{k}+\vec{q},\omega)
d\vec{k}$$, could give a reasonable count for the charge
modulations observed by STM\cite{Chatterjee,McElroy}. This joint
density-of-states describes the phase space for scattering of
electrons from the state at $\vec{k}$ to the state at
$\vec{k}+\vec{q}$ by certain modes with wavevector $\vec{q}$.
Therefore, one would expect that it peaks at the ordering
wavevector for $\omega=0$ near the phase transition of static
order. Since the states in the gapped Fermi patch are responsible
for the CDW here, autocorrelation analysis is conducted in the
normal state to study the CDW instabilities of $CDW^{65K}$ over
regions that would be gapped below $T_{CDW}$. The resulting
$C(\vec{q},0)$ is shown in Fig.\,\ref{correlation}\textbf{a} for
$\vec{q}$ along the $\Gamma$-M direction, which is mainly
consisted of a component at $\vec{q}=0$ that  exponentially
decays, and several features in Fig.\,\ref{correlation}\textbf{b},
where a peak is clearly observed around the CDW ordering
wavevector. The peak at $\vec{q}=0$ would require coupling to very
long wavelength phonons, which presumably is very weak.
Consistently, a recent calculation has shown that an optical
phonon branch softens significantly around \textbf{Q$_{i}$}, and
no sign of softening is observed at $\vec{q}=0$\cite{Lin}. In the
2D partial $C(\vec{q},0)$ map (obtained after deducting the
exponential decaying part and the constant background)in
Fig.\,\ref{correlation}\textbf{c}, although there are a few local
maxima corresponding to various possible orderings, the highest
peaks are those at \textbf{Q$_{i}$}.  Therefore, our results
suggest that the electronic structure is in favor of the charge
instability at \textbf{Q$_{i}$}'s, and eventually the system
becomes unstable to the CDW formation below $T_{CDW}$ in
collaboration with the phonons. Furthermore, it is also consistent
with the positive correlation between the spectral weight near
$E_F$ and T$_{CDW}$ in Fig.\,\ref{EDC}\textbf{i}.

The competition and coexistence of CDW and superconductivity can
be understood within the new framework. Recent photoemission
studies have revealed that superconducting gap opens at the
\textbf{K} and \textbf{$\Gamma$}
pockets\cite{shinScience,Valla2004}. The CDW gap opens at a
temperature higher than the superconducting transition, but it
just partially suppresses the density of states around the
\textbf{K} pocket and outer \textbf{$\Gamma$} pocket. Therefore,
as observed in most $2H$-TMD's, superconductivity is suppressed
but not eliminated by the CDW.

To summarize, the Fermi-patch mechanism of CDW in
2$H$-Na$_{x}$TaS$_{2}$ is characterized by the realization of both
ingredients of the CDW, energy gap and wavevector match on the
Fermi patches. Unlike other CDW mechanisms based on band structure
effects, it is rooted in the strong-coupling nature of its
electronic structure, which provides phase space needed for CDW
fluctuations. Furthermore, this mechanism would be realized not
only in polaronic systems, but also in materials where strong
electron correlations could cause Fermi patches, and thus CDW
instabilities.

We thank Profs. H. Q, Lin, J. L. Wang, Q. H. Wang, J. X. Li, Z. D.
Wang and F. C. Zhang for helpful discussions. This work is
supported by NSFC, MOST's 973 project: 2006CB601002 and
2006CB921300.


\begin{references}
\bibitem{exp3}The same $\alpha=0.71$ is used for the entire zone, and we have checked that
the gap size varies within $\pm 4\%$ against a $\pm 10\%$ change
of $\alpha$.
\end{references}
\end{document}